\pdfoutput=1
\documentclass[10pt,letterpaper]{article}
\usepackage{opex3}
\usepackage{amssymb,amsmath}
\usepackage{graphicx}
\usepackage{pstricks}

\begin{document}

\title{Large bandwidth, highly efficient optical gratings through high index materials}

\author{Helmut Rathgen$^{*1}$ and H. L. Offerhaus$^2$}

\address{$^1$Physics of Complex Fluids, University of Twente, The Netherlands.}
\address{$^2$Optical Sciences, University of Twente, The Netherlands.}
\email{helmut.rathgen@web.de}
\homepage{http://mrcwa.sourceforge.net/}

\begin{abstract}
We analyze the diffraction characteristics of dielectric gratings that
feature a high index grating layer, and devise, through rigorous
numerical calculations, large bandwidth, highly efficient, high
dispersion dielectric gratings in reflection, transmission, and
immersed transmission geometry.
A dielectric TIR grating is suggested, whose $-1dB$ spectral bandwidth
is doubled as compared to its all-glass equivalent. The short
wavelength diffraction efficiency is additionally improved by allowing
for slanted lamella. The grating surpasses a blazed gold grating over
the full octave. An immersed transmission grating is devised, whose
$-1dB$ bandwidth is tripled as compared to its all-glass equivalent,
and that surpasses an equivalent classical transmission grating
over nearly the full octave. 
A transmission grating in the classical scattering geometry is
suggested, that features a buried high index layer. 
This grating provides effectively $100$\% diffraction
efficiency at its design wavelegth, and surpasses an equivalent
fused silica grating over the full octave.
\end{abstract}

% ocis codes

% bibliography as generated through bibtex using osajnl.bst

% bibliography commands using bibtex
%\bibliographystyle{osajnl}
%\bibliography{strings,grating,computer,books,special}

\section{Introduction}

Diffraction gratings are an integral part of many modern optical
systems, with applications in lasers, imaging systems and
telecommunication. Recently {\em dielectric}
gratings have attracted increasing interest. Due to their high
resistance to laser induced damage paired with an unsurpassed
diffraction efficiency, etched fused silica gratings have found
widespread use in laser physics, for pulse compression
\cite{limpert02oe10_628,neauport05ao44_3143,clausnitzer03ao42_6934}
and wavelength control
\cite{nguyen97ol22_142}. On the other hand, 
the development of holographic replication techniques
has enabled large area volume gratings that are 
imprinted onto gelatine films sandwiched between two glass
plates, known as volume phase
holographic (VPH) gratings \cite{tamura05oe13_4125,rhee94ol19_1550}.
Their sealed layout (cleanability) and large possible dimensions
makes them the primary choice for the spectrographs of the recent
generation of astronomic telescopes
\cite{baldry04pasp116_403,ebizuka02procspie4842_319}.
Dielectric gratings are therefore of great interest.

Here, we introduce the use of high refractive index materials to
dielectric gratings.
Through rigorous numerical calculations, accompanied by intuitive
models, we investigate a number of new grating
designs -- representing total internal reflection types, immersed
transmission types, and classical transmission types --
that are superior to the state of the art, both in terms of peak
diffraction
efficiency and spectral bandwidth. Emphasis is put on devising designs
that are easily fabricated in practice.
The paper is organized as follows. In the remainder of this
introduction, we will walk through the principles of optical
gratings, and optical grating design, using the fused silica
TIR grating as an example. In Sec.\ref{sec:TIR}, we
present our numerical results for the diffraction
characteristics of dielectric TIR gratings based on high
index materials. In
Sec.\ref{sec:void}, we extend the use of high index materials to
immersed dielectric gratings. We first analyzing qualitatively the
challenge of designing an immersed grating that provides high
dispersion, and subsequently present our numerical results on immersed
gratings featuring a high index material. Finally, in
Sec.\ref{sec:trans}, the use of high index materials is extended to
classical dielectric transmission gratings. The paper is concluded in
Sec.\ref{sec:conclusions}.

\begin{figure}
  \centering
  \hspace{5mm}
  \begin{minipage}[b]{0.28\textwidth}
    \includegraphics[scale=0.42]{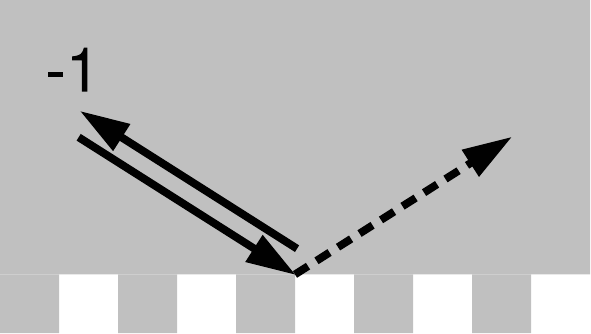}\rput(-30mm, 19mm){(a)}\\[1em]
    \includegraphics[scale=0.42]{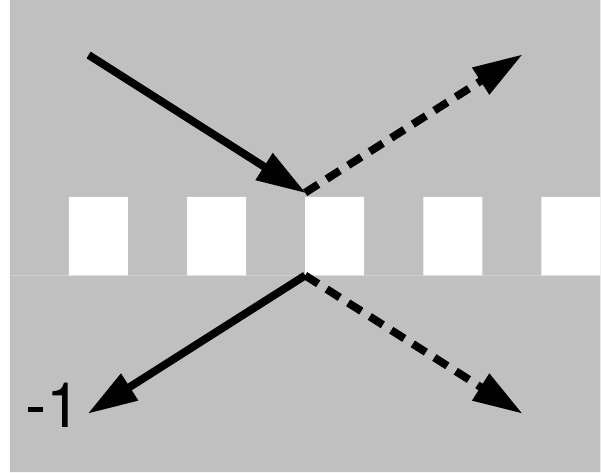}\rput(-30mm, 19mm){(b)}\\[1em]
    \includegraphics[scale=0.42]{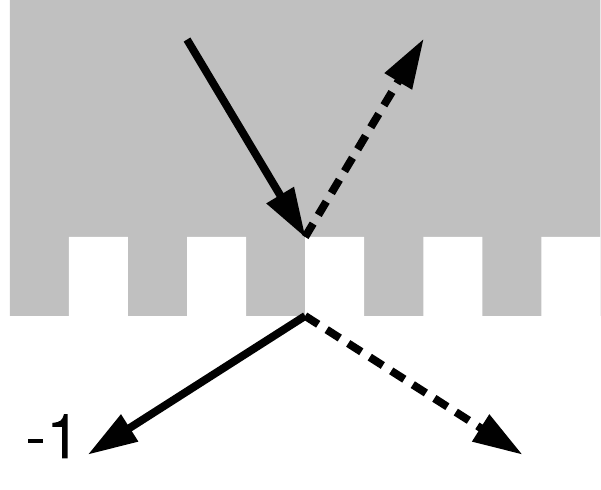}\rput(-30mm, 20mm){(c)}
  \end{minipage}
  \includegraphics[scale=0.6]{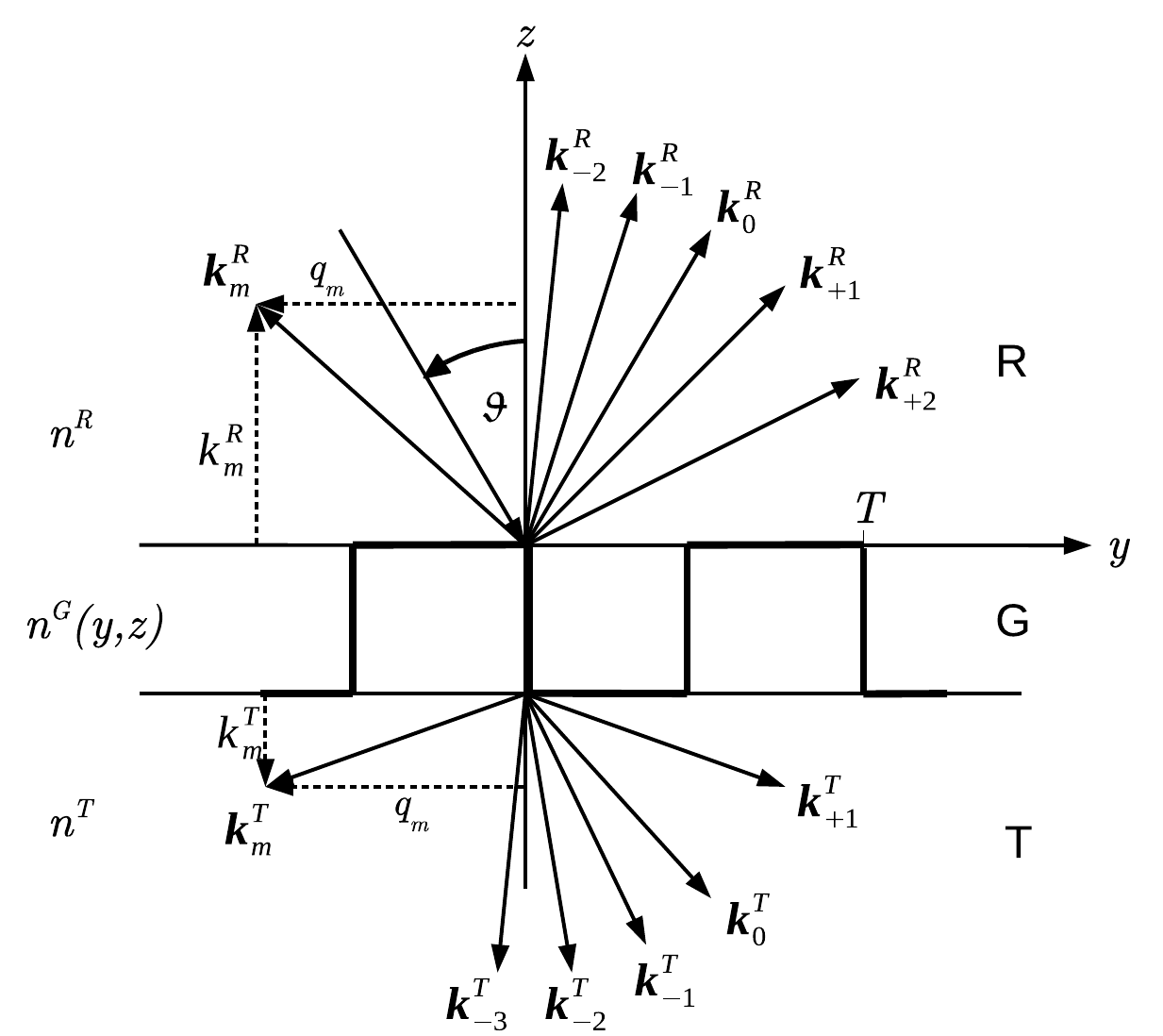}
  \rput(-72mm, 64mm){(d)}
  \caption{(a)--(c) Schematics of the investigated types of dielectric
    gratings, (a) TIR grating, (b) immersed grating, (c) classical
    transmission grating. (d) Scattering geometry of an optical grating.}
  \label{fig:geometry}
\end{figure}

If a dielectric grating is illuminated under a
sufficiently large incident angle, total internal reflection (TIR)
occurs at the grating surface, and the device is turned from a
transmission grating into a reflection grating. This configuration
offers great benefits. Most importantly, reflection is achieved
without the need of a metallic coating, and thus, the low absorption
loss of the grating, and consequently a large diffraction
efficiency, paired with a large resistance
against laser induced damage is maintained. Given the more than 350
year long history of the optical grating \cite{gregory}, it is
indeed remarkable, that this scattering geometry was discovered only
recently
\cite{maricante04ol29_542,liu07optcom273_290,livingston96ieee46_958}.

Fig.\ref{fig:geometry}(a) shows the schematic of the dielectric
TIR grating. Similar to an ordinary transmission grating,
rectangular grooves are fabricated into the backside of a glass
body, and the grating is illuminated from the glass side. However,
in contrast to an ordinary transmission grating, the grating is
illuminated under an incident angle that is larger than the angle
of total internal reflection between glass and air ($\vartheta_T
\approx 43.6^\circ$), such that the zero-order transmitted beam is
evanescent. Furthermore, the period $T$ of the grating is chosen
sufficiently small, such that all other diffraction orders on the
transmission side are evanescent as well. This is achieved by
stressing the scattering geometry of the grating.

Fig.\ref{fig:geometry}(d) illustrates the scattering geometry.
Generally, as an optical grating is illuminated with a plane wave,
a number of plain waves are scattered from the grating. Waves are
scattered into the reflection region {\sf R} , and into the
transmission region {\sf T}, that are characterized each by their
refractive index $n^R$, respectively $n^T$. Associated with every
scattered wave is an amplitude and a wavevector. The wavevector
expresses the direction of the wave -- or, in case the wavevector
is complex, the penetration depth of the evanescent wave. The
components of the wavevectors in the periodic direction (here the
$y$-components, see Fig.\ref{fig:geometry}) are determined by the
Bloch condition as $q_m=q_0+mQ$, where $m=-\infty,\dots,+\infty$
is the diffraction order, $Q=\lambda/T$, and
$q_0=n^R\sin\vartheta_0$. Here $\vartheta_0$ is the incident
angle. We have non-dimensionalized the spatial coordinate through
$r\mapsto 2 \pi r / \lambda$. The $z$-components of the
wavevectors are determined through the wave equation as $k_m^S =
({n^S}^2-{q_m^S}^2)^{1/2}$. $S=R,T$ denotes the reflection and
transmission region, respectively. A real $k$ represents a
propagating diffraction order, a complex $k$ represents an
evanescent wave. The latter relations imply the grating equation
\begin{equation}\label{grating}
  \sin\vartheta^S_m=\sin\vartheta^S_0+\frac{m\lambda}{n^ST},
\end{equation}
which expresses the angles of the scattered waves.
Angles are measured with a positive sign in positive $y$-direction,
and with a negative sign in negative $y$-direction.
The angular dispersion $D$ follows by differentiating the latter equation
with respect to $\lambda$,
\begin{equation}\label{dispersion-general}
  D=\frac{\partial \vartheta^R_m}{\partial \lambda} = -\frac{m}{Tn^R\cos\vartheta^R_m}.
\end{equation}

The TIR grating is then arrived at, by requiring
\begin{equation}\label{TIR}
  |q_m|>n^T.
\end{equation}
This determines a range, where the period of the grating
is sufficiently small.

Generally, the highest diffraction efficiency is achieved if the
grating is mounted in $-1^{st}$ order Littrow configuration, where
the $-1^{st}$ diffraction order is antiparallel to the incident
wave, that is $-q_{-1}=q_0$. To see the benefit of this, consider
the opposite case: Suppose a binary (rectangular profile)
reflection grating is illuminated at zero incident angle, that is,
perpendicular to the surface. And suppose the period is small
enough, such that only the $+1^{st}$ and the $-1^{st}$ diffraction
order are present. It is possible to choose a suitable width and
depth of the grating grooves, such that the largest part of the
diffracted intensity is scattered into the $0^{th}$ order, or, by
choosing a different width and depth, all of the diffracted
intensity is scattered into the first order. However, due to
symmetry, the intensity of the $+1^{st}$ and the $-1^{st}$
diffraction order must be equal, such that, at maximum 50\% of the
diffracted intensity is scattered into a single diffraction order.
In contrast, if the incident angle is chosen such that the
$-1^{st}$ diffraction order is antiparallel to the $0^{th}$ order,
only the $0^{th}$ and the $-1^{st}$ order are present, and the
scattered intensity can be distributed at will among those two
orders (by choosing a suitable groove width and depth), thus
(nearly) 100\% diffraction efficiency can be achieved. Remarkably,
near 100\% diffraction efficiency can be achieved also with higher
order Littrow configurations, characterized by $-q_m=q_0$, and
even more surprising, this holds true even for higher order TIR
gratings, where no longer all transmitted orders are forbidden
(see supplemental Fig.1). This implies indeed, that the TIR
grating condition can be relaxed. Rather than requiring that {\em
all} diffraction orders on the transmission side be evanescent (as
suggested originally in Ref.\cite{maricante04ol29_542}, and
represented by Eq.\ref{TIR}), it is sufficient to require that the
$0^{th}$ transmitted order is evanescent, i.e., the ordinary
condition for total internal reflection is satisfied, expressed as
$|q_0|>n^T$. Independent of that, in most cases, one will design a
grating for $-1^{st}$ order Littrow configuration, because here,
the overlap between higher diffraction orders is minimal -- the
angular range per diffraction order is largest -- such that the
largest spectral bandwidth can be achieved.

The Littrow configuration settles the period of the grating as
\begin{equation}\label{period}
  T=m\lambda/(2n^R\sin\vartheta^R_0).
\end{equation}
From now on, we assume the wavelength is $1064$nm, the refractive
index of the glass is $1.45$, and the incident angle is
$60^\circ$. Thus, the period of the grating is $T=423.66$nm in our
case. This settles also the angular dispersion, since
the Littrow condition substituted in Eq.\ref{dispersion-general} yields
\begin{equation}\label{dispersion}
  D=2/\lambda\,\tan\vartheta_0.
\end{equation}
The dispersion is minimal at zero angle (perpendicular to the
surface), and diverges at grazing angles.
With the above parameters, the dispersion is $D=0.187^\circ/$nm. A comparably
large dispersion is inherent to TIR gratings, because of the
necessarily large incident angle.

To arrive at a highly efficient TIR grating, one varies the width
$w$ and depth $d$ of the grating grooves, and evaluates the
resulting intensity of the $-1^{st}$ diffraction order with a
suitable numerical method. A number of efficient numerical methods
have been developed over the past decades, that allow to do so.
Here, we use the multilayer rigorous coupled wave analysis, as
devised in Ref.\cite{moharam95josaa12_1077}.  The numerical code
used in this work \cite{rathgen07mrcwa}, is written in FORTRAN\ 90
and Python \cite{python}, and is made freely available under the
GPLv3; we aim to provide a computational tool to the community,
that can be openly reviewed (open source) and is free of charge. A
flexible and highly intuitive user interface is provided through a
set of python bindings, that are easily extended and embedded into
complex computational tasks. The numerical core routines are
provided by optimized FORTRAN\ 90 code, relying heavily on the
high performance linear algebra package LAPACK
\cite{anderson99lapack_users_giude}, typically in form of the
intel\texttrademark MKL. In this work, all calculations were
performed on a grid equipped with 2GHz Dual Core AMD
Opteron\texttrademark\ processors, and took only few hundred cpu
hours.

\begin{figure}
  \centering
  \hspace{-4mm}
  \includegraphics[scale=0.4]{icon-TIR.pdf}\rput(-24mm, 16mm){(a)}
  \hspace{1.2em}
  \includegraphics[scale=0.4]{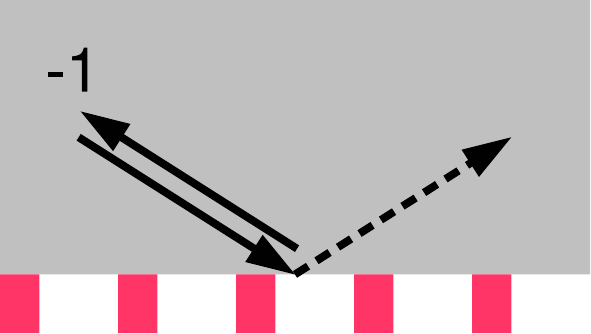}\rput(-24mm, 16mm){(b)}
  \hspace{1.2em}
  \includegraphics[scale=0.4]{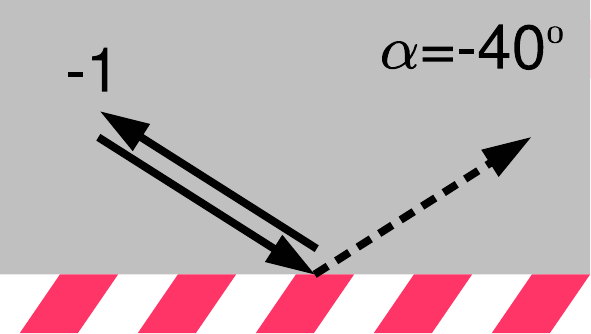}\rput(-24mm, 16mm){(c)}
  \hspace{1.2em}
  \includegraphics[scale=0.4]{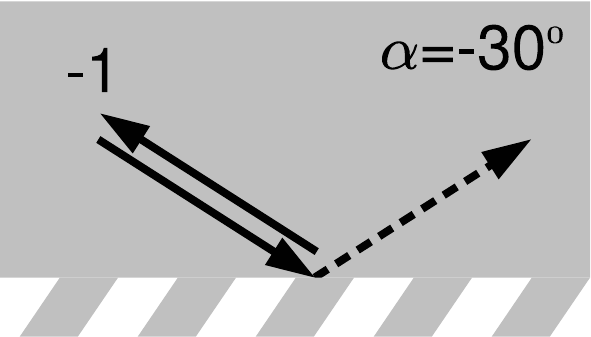}\rput(-24mm, 16mm){(d)}\\
  \includegraphics[scale=1]{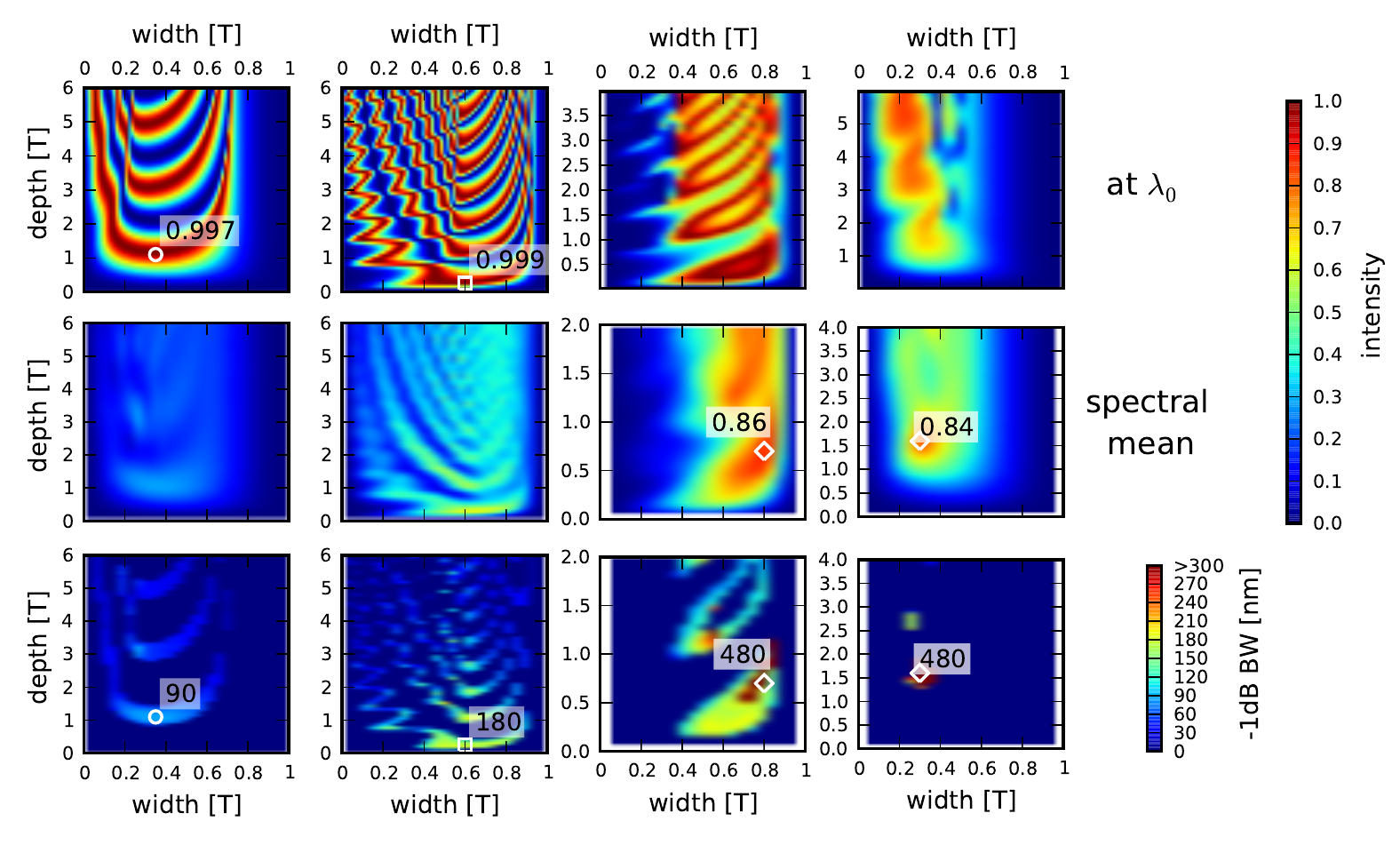}
  \caption{Diffraction efficiency of dielectric TIR
    gratings as a function of the groove width
    and depth. (a)
    silica TIR grating. (b) TIR grating based on a high index
    material (n=2.4). (c) TIR grating based on a high index material
    featuring slanted lamella ($\alpha=40^\circ$). (d) silica TIR
    grating featuring slanted lamella ($\alpha=30^\circ$). The upper
    panels show the diffraction efficiency at
    the design wavelength $\lambda_0=1064$nm. The middle panels show
    the mean diffraction efficiency, averaged over the theoretically
    accessible spectral band, as determined by the condition of
    non-overlapping diffraction orders,
    $\lambda_{cut}/2 < \lambda < \lambda_{cut}$, where
    $\lambda_{cut}=n^RT(1+\sin\vartheta_0)$. The lower panel shows
    the $-1dB$ spectral bandwidth. The open white symbols mark
    optimal choices of the parameters $w$ and $d$ for each type of
    grating, and represent those values, for which the spectral
    characteristics is plotted in Fig.\ref{fig:TIR-lambda}. All
    results shown correspond to $s$-polarized light.}
  \label{fig:TIR-w-d}
\end{figure}

The upper panel of Fig.\ref{fig:TIR-w-d}(a) shows the calculated
diffraction intensity of the $-1^{st}$ diffraction order as a
function of groove width and depth, for $s$-polarized light
(results for $p$-polarization and averaged $s$ and $p$ are
included in the supplemental material). Broad regions appear,
where the diffracted intensity is close to unity. This behavior
can be understood by considering the propagation constants in
normal direction, of the modes that are excited inside the grating
region \cite{clausnitzer05oe13_10448}. As the grating is arranged
in $-1^{st}$ order Littrow configuration, only the fundamental and
the first mode in the grating region carry a notable amount of
energy. They propagate at different velocities, determined by
their propagation constants. Thus, as they propagate, they
accumulate a phase shift. The phase shift that the waves have
accumulated as they couple out of the grating region,
determines whether they are scattered into the ordinary reflected
beam or into the $-1^{st}$ order. In this way, by choosing the
depth of the grooves, the intensity can be distributed at will
among the $0^{th}$ and the $-1^{st}$ order. The propagation
constants of the modes are determined by their effective
refractive indices. Those depend on the filling fraction of the
grating, that is, the groove width. Consequently, depending on the
groove width, a different depth is required to achieve the desired
phase difference. The fringes in Fig.\ref{fig:TIR-w-d}(a)
correspond to integer multiples of the interference condition. It
should be noted, that, along a line in the center of each fringe,
the intensity is truly unity. This is in contrast to classical
transmission gratings, where the maximum theoretical intensity is
limited by the effective reflectivity of the grating, which
inevitably results in a scattering loss into the $0^{th}$ order,
such that 100\% diffraction efficiency cannot be reached (e.g.
\cite{clausnitzer08oe16_5577}, and the discussion in
Sec.\ref{sec:void}, and Fig.\ref{fig:trans-w-d}(a)). The latter
elevates the TIR grating as a unique component for applications
where highest diffraction efficiencies are required. The
width-depth-map suggests a number of possible choices for the
width and depth, that result in a highly efficient TIR grating. To
devise a TIR grating that is easily fabricated in practice, one
may chose an intermediate width and a small depth, corresponding
to a small aspect ratio. The white circle in the upper panel of
Fig.\ref{fig:TIR-w-d}(a) represents one possible choice, with
$w=0.35T$ and $d=1.1T$. To find the spectral characteristics of
the corresponding grating, one varies the wavelength of the
incident light, and evaluates the resulting diffraction
efficiency. The latter may be done in two ways. (1) The incident
angle is held fixed, while the wavelength is varied -- this case
is encountered, e.g., with gratings that are used for dispersion
control in femtosecond laser applications, or with spectrometers
that are equipped with a line CCD and operate without movable
parts. (2) As the wavelength is varied, the incident angle is
adapted, such that the grating continues to satisfy the Littrow
condition for every wavelength -- this case is encountered, e.g.,
in a classical scanning monochromator. Here, we consider the first
case, motivated by the application of the dielectric TIR grating
for pulse compression, and furthermore, because the range of
possible incident angles for the TIR grating is not particularly
large.

\begin{figure}
  \centering
  \includegraphics[scale=1]{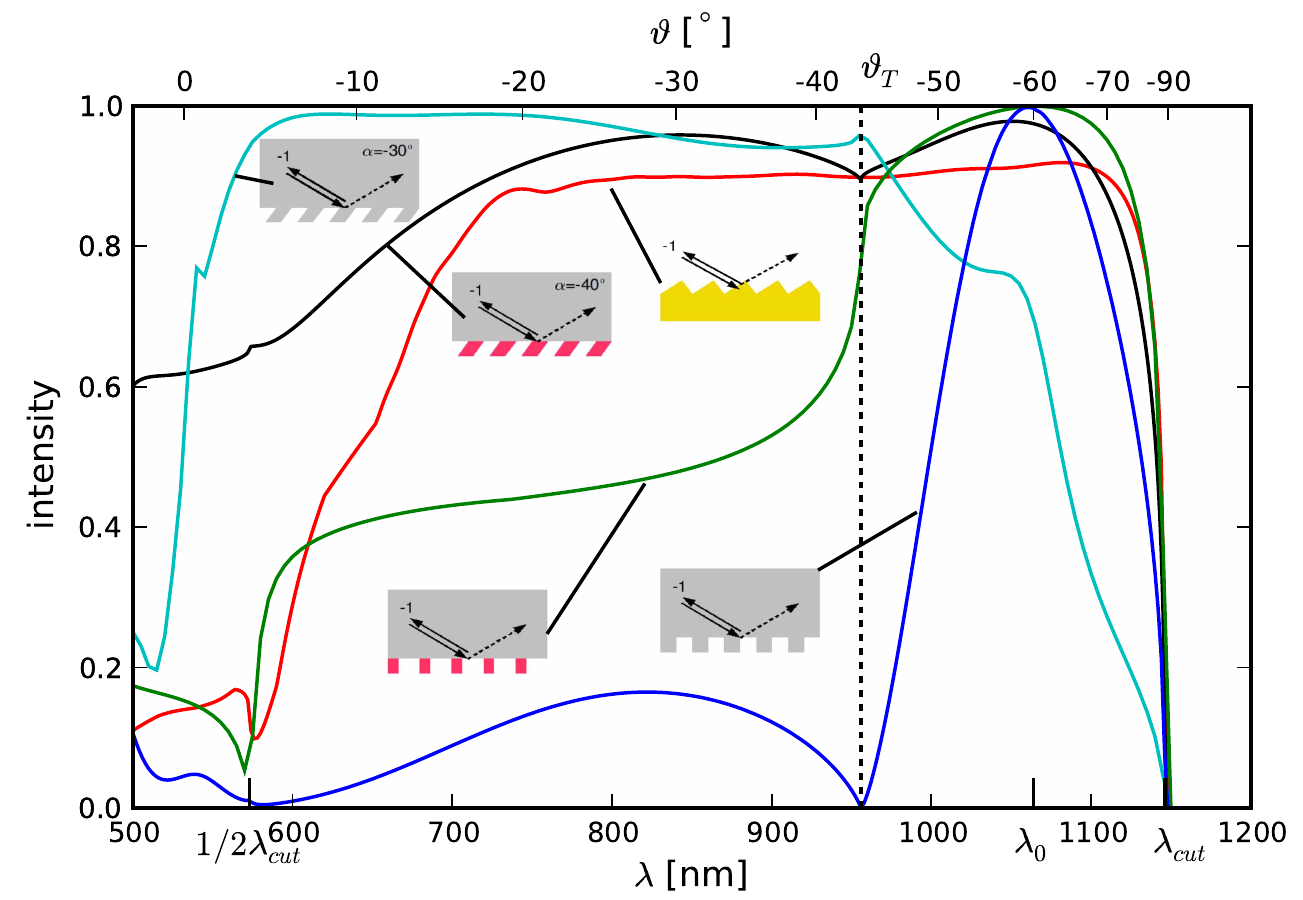}
  \caption{Spectral characteristics of
    dielectric TIR gratings ($s$-polarized light), as compared
    to a best blazed gold grating with blaze angle $33^\circ$
    ($p$-polarized light).
    The blue line shows the result for a silica TIR grating with
    groove width $0.35T$ and depth $1.1T$, corresponding to maximum
    peak efficiency and spectral bandwidth (white circle in
    Fig.\ref{fig:TIR-w-d}(a)). The green
    line shows the spectral characteristics of a dielectric TIR
    grating featuring a high index material
    ($n=2.4$) in the grating region
    ($w=0.6T$, $d=0.25T$, as marked by the white square in
    Fig.\ref{fig:TIR-w-d}(b)). The
    black line shows the spectral characteristics of
    a dielectric TIR grating with a high index material and
    slanted lamella ($\alpha=40^\circ$), groove width $0.7T$, depth $0.5T$,
    corresponding to an optimal choice
    for large mean diffraction efficiency and simultaneously
    a comparably smooth spectrum, (white diamond in
    Fig.\ref{fig:TIR-w-d}(c)). The cyan line shows the result for a
    silica TIR grating with slanted lamella ($\alpha=30^\circ$),
    groove width $0.3T$, depth $1.6T$, optimized for
    high short wavelengths efficiency.}
  \label{fig:TIR-lambda}
\end{figure}

The blue line in Fig.\ref{fig:TIR-lambda} shows the spectral
characteristics of the dielectric TIR grating, with width $0.35T$
and depth $1.1T$, in comparison to a best blazed gold grating (in
air), with period $T=614$nm, that has the same dispersion and
incident angle as the TIR grating. The blaze angle of the gold
grating was optimized to yield the largest possible bandwidth and
peak efficiency in the given wavelength range, found as
$33^\circ$. The dielectric TIR grating shows excellent properties
in a 50-100nm band around its design wavelength. In particular, it
reaches 100\% diffraction efficiency at its design wavelength, and
provides scattering losses smaller than -1dB (95\% diffraction
efficiency) over a bandwidth 
of 84nm. Here, it provides unsurpassed diffraction efficiency.
Indeed, these high diffraction efficiencies cannot be reached with
metallic gratings, whose inherent absorption inevitably results in
losses \cite{petit80electromagnetic_theory_of_gratings}. However,
as the wavelength is altered substantially from the
design wavelength, the diffraction efficiency quickly drops to
zero. In particular, as the angle of the $-1^{st}$ diffraction
order reaches the angle of total internal reflection between glass
and air, the diffraction intensity drops to a sharp minimum, and
for shorter wavelengths -- corresponding to angles smaller than
the angle of total internal reflection -- the diffraction
intensity does not recover its initially large value. These
diffraction angles correspond to the case, when the $-1^{st}$
diffraction order on the transmission side is no longer
evanescent. However, a large diffraction intensity of the
reflected $-1^{st}$ order is not strictly forbidden at these
scattering angles, as can be seen by the small but nonzero
intensity value. This low diffraction efficiency over the largest
part of the spectrum renders the TIR grating unusable for a large
number of applications, including broadband pulse compression and most
spectroscopic applications.

It is desirable, to resolve this issue, and to increase the
spectral bandwidth of the TIR grating towards values that are
comparable with metallic reflection gratings and dielectric
transmission gratings, while maintaining the extraordinarily large
peak diffraction efficiency, and the high resistance to laser
induced damage. This is the first objective of this paper.

\section{Large bandwidth dielectric TIR grating}\label{sec:TIR}

The described phenomenon of a suppressed diffraction intensity at
smaller angles, is remarkably robust. A first approach to increase
the spectral bandwidth would be, to explore the landscape of
possible combinations of the width and the depth, and to evaluate
how this effects the spectral characteristics. However, the
spectral bandwidth is only decreased. The lower panel of
Fig.\ref{fig:TIR-w-d}(a) shows the $-1dB$ spectral bandwidth as a
function of the grove width and depth. It is seen, that the
previous choice (with $w=0.35T$ and $d=1.1T$) corresponds already
to the largest possible bandwidth, and for larger depths, the
reflectivity and the bandwidth decrease in an oscillatory fashion.
This behavior is understood qualitatively by referring to the theory
of Kogelnik, who gave an estimate for the bandwidth $\Delta\lambda$
(full width at half maximum) of dielectric transmission gratings in
the limit of deep grooves as
\begin{equation}
  \frac{\Delta\lambda}{\lambda}=\frac{T}{d}\cot \vartheta_0.
\end{equation}
It is seen, that indeed, the bandwidth is largest at small $d$.

This qualitative insight suggests already a way to improve the
spectral characteristics of the grating. As an increase of the
groove depth reduces the spectral bandwidth of the grating, one
should try the opposite, one should decrease the groove depth. To
do so, the difference of the propagation constants of the two
modes in the grating region has to be enlarged, that is, the
refractive index contrast has to be increased. Thus, in the
grating region, we replace the glass by a material that has a
larger refractive index. Suitable high index optical materials
that are routinely available in the form of high quality thin
films -- through optical coating applications -- are, e.g.,
Sapphire, diamond, TiO$_2$, etc.. The latter two materials offer
the largest refractive index, around 2.4, in the wavelength range
considered here \cite{palik85handbook_optical_constants}. From now
on, we assume the refractive index of the high index material is
2.4. It should be noted, that the glass
in the reflection region is kept. This is indeed important. If the
material in the reflection
region was changed together with the groove material, the grating
period needed to be decreased accordingly. As a result, the entire
computational problem was rescaled, such that no improvement was
achieved.

The upper panel of Fig.\ref{fig:TIR-w-d}(b) shows the diffraction
efficiency as a function of groove width and depth for the high index
TIR grating. It is seen, that indeed, the first fringe is shifted
towards shallower
grooves. The smallest possible depth is achieved at around $w=0.6T$
and $d=0.25T$ (white square in
Fig.\ref{fig:TIR-w-d}(b)). And indeed, here, the spectral bandwidth
reaches its maximum (lower panel of Fig.\ref{fig:TIR-w-d}(b)).
The green line in Fig.\ref{fig:TIR-lambda} shows the corresponding
spectral characteristics. The diffraction
efficiency is increased over the entire spectrum as compared to the
TIR grating based on glass. In particular,
the $-1dB$ bandwidth around the design
wavelength, is increased to 180nm, and is more than doubled as compared to
the all-in-glass equivalent.
Additionally, towards smaller wavelengths, in the
range where the diffraction angle is larger than the angle of total
internal reflection between glass and air, the diffraction efficiency is
greater than 40\%. This is a useful value for many spectroscopic
applications. In addition, the moderate aspect ratio of the grating
grooves ($\approx 0.5$), suggests that such a grating could be fabricted
rather easily. Furthermore, the different nature of the grating and
substrate material will probably help to fabricate structures with a
very well defined depth, determined only by the thickness of the layer
that is deposited prior to the etching process. The thickness of
deposited layers can typically be controlled very accurately.
Nevertheless, the natural question arises, whether this grating design
can be further improved.

We have so far considered only the most simple of all grating
profiles -- rectangular grooves. It is well known that the
diffraction efficiency of a gold grating -- in particular its
spectral characteristics -- can be greatly improved, by choosing
an appropriate blaze angle
\cite{petit80electromagnetic_theory_of_gratings}. The blaze angle
is chosen such that the incident beam and the diffracted beam are
nearly perpendicular to the long face of the grating tooth (over
the range of incident angles determined by the desired spectral
range). Could the spectral characteristics of the TIR grating be
further improved by using an equivalent geometry? As we will show
below, the answer is yes, however, it is not sufficient to
introduce a blaze, which becomes clear by considering the depth of
the grooves necessary to accumulate the required phase shift in
the grating region. As the geometry is more complex, the
propagation constants of the waves vary in $z$-direction over the
grating region, nevertheless, the mechanism for achieving maximum
diffraction efficiency through the constructive interference of
the fundamental and the first mode, is still qualitatively the
same. And indeed, near 100\% diffraction efficiency can be
achieved with a large variety of groove geometries, such as
symmetric triangles, asymmetric triangles, bow-ties, etc., by
choosing a suitable width and depth. However, as is the case with
rectangular grooves, generally a relatively large total depth is
required, since the difference of the effective index of the
propagation constants is never very large. E.g, in the case of
symmetric triangles, a height of several $T$ is required to
achieve near 100\% diffraction efficiency. In contrast, the blazed
geometry is characterized by a small aspect ratio. In particular,
the aspect ratio is smaller than $T/2$ (the blazed geometry is
characterized by a triangle that features a right angle in its
tip, related to the original ruling process that uses an inclined
rectangular cutter). Because of this small aspect ratio, the
blazed geometry cannot provide a high diffraction efficiency, for
the type of dielectric gratings discussed here. However, a
geometry that provides inclined faces, without sacrificing the
possibility of a large groove depth, is the slanted geometry,
characterized by rectangular lamella that are sheared along the
periodic direction with a slant angle $\alpha$. The slant angle is
defined as the angle by which the groove wall is rotated as
compared to the normal direction. Clockwise rotation corresponds
to positive angles. Indeed, the slanted geometry has much in
common with the geometry of volume phase holographic gratings,
where, in the general case, the crests of the holographically
imprinted refractive index modulation are allowed to be slanted
with respect to the grating normal (e.g.
\cite{baldry04pasp116_403}).

To evaluate the diffraction characteristics of a slanted TIR
grating, we compute the diffraction efficiency as a function of
groove width, depth and wavelength for all slant angles
$-90^\circ<\alpha<90^\circ$, and for each slant angle, we evaluate
the spectral bandwidth. The optimum slant angle is found as
$\alpha=40^\circ$. Fig.\ref{fig:TIR-w-d}(c) shows the plot (a
movie sequence showing the plots for all $\alpha$, is included in
the supplemental material). The $-1dB$ bandwidth exceeds 300nm in
a region around $w=0.8T$, $d=0.7T$. The black line in
Fig.\ref{fig:TIR-lambda} shows the corresponding spectral
characteristics. This grating has a diffraction efficiency larger
than 90\% over nearly the full octave, and outperforms the best
blazed gold grating over the largest part of the spectrum. In
particular, two broad peaks of high diffraction efficiency appear,
at around $30^\circ$, and at the design angle, $60^\circ$. These
broad peaks overlap, and provide very high diffraction efficiency
over the hole spectrum. Around each of the peaks, the spectrum is
smooth. A shallow kink remains at the wavelength that
corresponds to a diffraction angle equal to the angle of total
internal reflection. This kink is at the origin of the jump that
is observed in the width-depth map of the bandwidth, at the boundary
of the region of large bandwidth. As the kink reaches above the
$-1dB$ limit, the bandwidth jumps from a value characterized by
the width of the right peak, to a value characterized
by the width of both peaks together. It should be
noted, that a combination of width and depth can be chosen, such
that the kink between the two peaks at the angle of total internal
reflection, disappears, and the spectrum becomes completely smooth
over the full spectral range. However, in this case the spectral
bandwidth is slightly smaller. Similarly, a width and depth can be
chosen such that the right peak maintains 100\%
diffraction efficiency at the design wavelength, while the left
peak is still higher than 80\%. However, in this case, the kink
becomes more pronounced. The result plotted here, represents a
compromise between largest possible bandwidth and smoothness of
the kink. Independent of that, it may also be interesting to
operate this grating only with scattering angles smaller than the
angle of total internal reflection, that is, the incident light
enters the grating under a large angle, and the scattered light
returns under angles smaller than the angle of total internal
reflection. Thereby, only the left part of the spectrum was used.
The design could then be optimized to yield highest efficiency
specifically across the left peak.

The great improvement of the spectral characteristics provided by
slanted lamella raises the question whether slanted lamella would
also improve a single component TIR grating fabricated entirely
from glass. As we will show now, the answer is yes, and it turns
out that this is only achieved by optimizing the design in the way
just described above, that is, by utiltizing the left peak of the
spectrum, while the right peak, at the design wavelength, remains
small. This is indeed peculiar, because it means, that the grating
does {\em not} feature high efficiency when the wavelength
satisfies the Littrow condition, but at another wavelength, which
also means, that one overlooks a highly efficient design, if one
optimizes the grating parameters in the usual way at the design
wavelength (because 
there the intensity is small). This is indeed remarkable. Recall
the intuitive arguments that we put forward to explain why the
Littrow configuration leads to a high efficiency. We considered
the opposite case, when the incident wave is perpendicular to the
surface, and the $+1^{st}$ and the $-1^{st}$ diffraction order are
aligned symmetrically around the surface normal. We argued, that,
in this case, due to symmetry, the intensity of the orders must be
equal, such that at maximum 50\% diffraction efficiency can be
achieved, and in contrast, if the grating is arranged in Littrow
configuration, (near) 100\% diffraction efficiency can be
achieved. However, with the slanted geometry, the mirror symmetry
of the grating around the surface normal is broken, and these
arguments are no longer rigorously valid. It turns out that, for
the slanted TIR grating in silica, this is indeed strongly
violated, and the Littrow configuration is no longer the universal
optimum scattering geometry, rather, the optimum is achieved when
the white light enters under a large angle, and the arc of
diffracted beams leaves under small angles. To the best of our
knowledge, an efficient grating with this type of scattering geometry
has not been reported before.

Fig.\ref{fig:TIR-w-d}(d) shows the diffraction efficiency as a
function of groove width and depth for the slanted silica TIR
grating. As with the slanted high index TIR grating, we have
evaluated the dependence of the diffraction characteristics on the
slant angle (see second movie in the supplemental material), and
present here the optimum, found with $\alpha=30^\circ$. The
largest bandwidth is found with width $w=0.3T$ and depth $d=1.6T$.
The corresponding spectral characteristics are shown with a cyan
line in Fig.\ref{fig:TIR-lambda}. The grating has excellent
spectral characteristics for wavelength corresponding to
diffraction angles below the angle of total internal reflection.
For larger angles, the intensity falls off quickly, and at the
design wavelength, it has already dropped below the $-1dB$ limit.
This remarkable scattering geometry provides an unsurpassed
diffraction efficiency over an almost 500nm band towards the short
wavelength side of the spectrum. To the best of our knowledge,
this represents again, the largest bandwidth highest efficiency
reflection type optical grating devised so far.

\section{Large bandwidth immersed grating}\label{sec:void}

The described dramatic improvement of the spectral bandwidth of a
dielectric TIR grating through a high index material, is more
general. In particular, the introduction of a high index material
improves dramatically also the performance of immersed dielectric
gratings \cite{nishii04ao43_1327} at large incident angles. This is
important, since, by
virtue of Eq.\ref{dispersion}, large angles correspond to large
dispersion. In addition, the immersed dielectric grating has been
suggested previously, for providing a device that is easily
cleaned, and usable in a rough (dirty) environment.
Fig.\ref{fig:geometry}(b) shows the schematic of an immersed
dielectric grating. The system is identical to the TIR grating,
except that the air on the transmission side is replaced by a
glass body. Besides the practical advantages related
to cleaning, immersed dielectric gratings have another particular
advantage over ordinary transmission gratings. In contrast to the
latter, they can provide theoretically 100\% diffraction
efficiency. The latter is the topic of a recent paper
\cite{clausnitzer08oe16_5577}, where the authors show, that 100\%
diffraction efficiency is linked to equal effective reflectivity
of the top and bottom boundary of the grating region. If the
incident and transmission halfspaces have a different refractive
index -- as is the case for an ordinary transmission grating,
where the transmission region is air -- the reflectivities of the
top and bottom grating boundary are different, such that, even if
an optimal interference condition is met for the two modes that
travel in the grating region, a finite intensity remains, that
results in an inevitable reflection loss into the ordinary
reflected order. Thus, the diffraction efficiency is at best one
minus an effective reflectivity of the optical grating. The latter
can be substantial, especially when highest efficiencies are
required, as is the case in many laser applications, and
especially with large angles, where the grating provides its
largest dispersion. In contrast, the immersed dielectric grating
does not suffer this deficiency, and it can theoretically provide
100\% diffraction efficiency -- analog to the TIR grating.

However, as was the case with the TIR grating, the spectral
bandwidth is limited, and, more importantly, if operated with
large incident angles
-- corresponding to large dispersion --, the immersed grating
starts to loose its extraordinarily good characteristics.

\begin{figure}
  \centering
  \begin{tabular}{cc}
    \rput(-18mm,16mm){(a)}
    \includegraphics[scale=0.3]{icon-void.pdf}&
    \rput(-18mm,16mm){(b)}
    \includegraphics[scale=0.3]{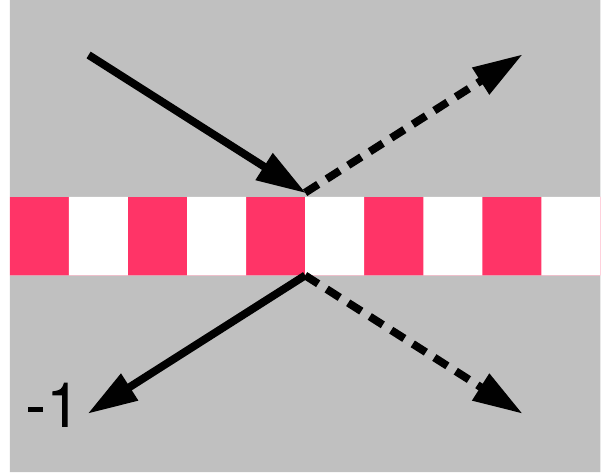}\\
    \includegraphics[scale=0.45]{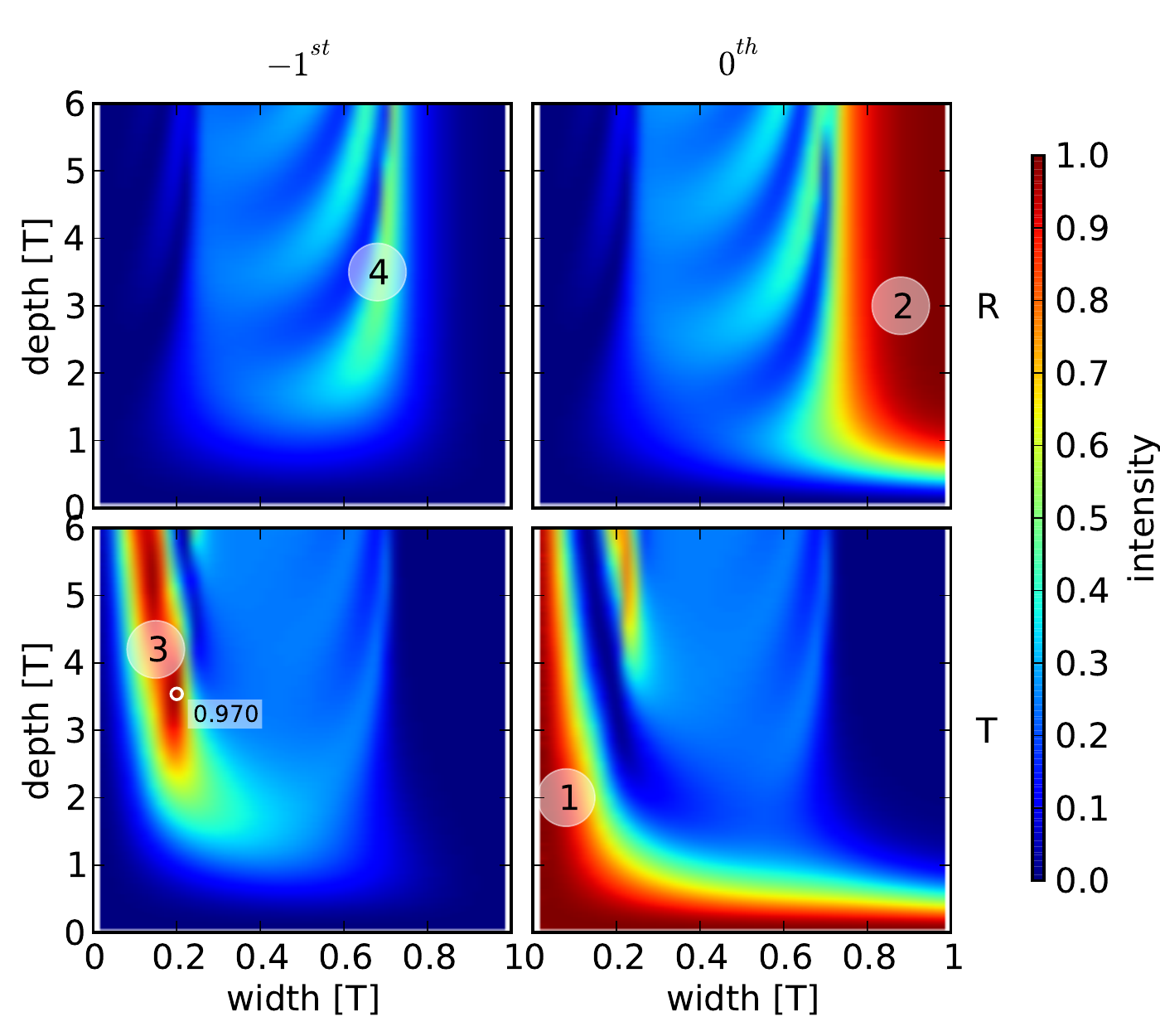}&
    \includegraphics[scale=0.45]{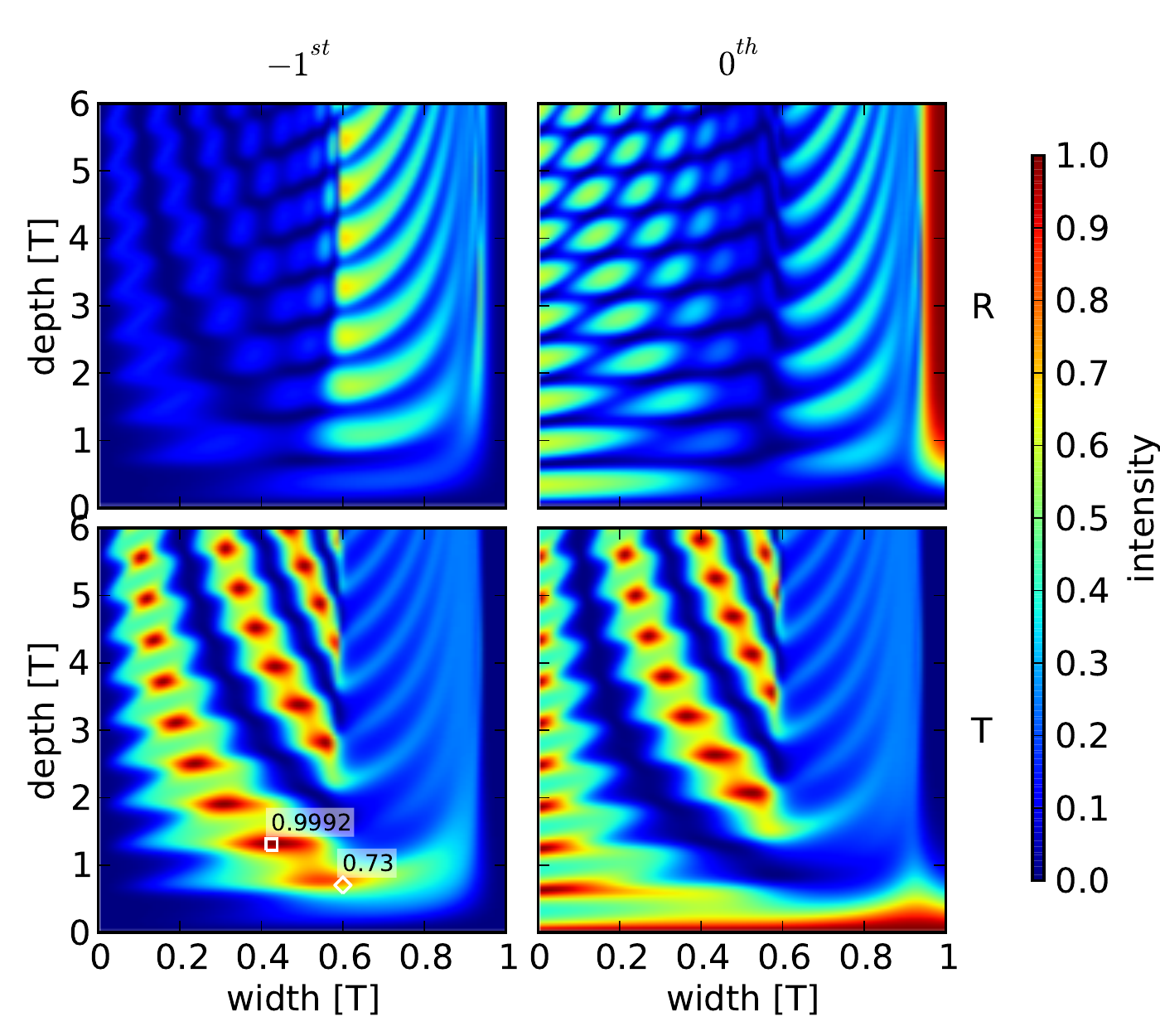}\\
  \end{tabular}
  \caption{(a) diffraction efficiency for an immersed dielectric
    grating, illuminated under a large incident angle
    ($\vartheta_0=60^\circ$). The upper panels show the reflected
    orders, the lower panels show the transmitted orders. The left and
    right panels show the $-1^{st}$ and $0^{th}$ order,
    respectively. (b) same as in (a), for a grating featuring a high
    index material ($n=2.4$) in the grating region.}
  \label{fig:void-w-d}
\end{figure}

The four left panels of Fig.\ref{fig:void-w-d} show the
diffraction efficiency ($s$-polarization) as a function of groove
width and depth, for a classical immersed dielectric grating
illuminated under a large incident angle (throughout this section, we
consider the same incident angle, $\vartheta_0=60^\circ$, and period,
$T=423.66$nm, as before, linked to the $-1^{st}$ order Littrow
condition, and providing the same dispersion). Four regimes
appear. (1) At small groove depth, the ordinary transmitted
intensity is large -- the grating is negligible. (2) At large
groove depth and large groove width, the ordinary reflected
intensity is large -- the grating region is increasingly similar
to a large air gap, and total internal reflection occurs at the
top grating surface. (3) Towards smaller groove widths, a region
exists, where nearly all of the scattered intensity is scattered
into the transmitted $-1^{st}$ order -- the grating resembles an
efficient transmission grating. (4) Towards larger groove widths,
a region exists, where a large part of the scattered intensity is
scattered into the reflected $-1^{st}$ order -- the grating
resembles a TIR grating. Thus, when illuminated with a large
incident angle, depending on the groove width and depth, the
immersed dielectric grating can be either more akin to a TIR
grating or more akin to a transmission grating or somewhere in
between. As a consequence of that, the regions of optimal choice
of width and depth shrink to narrow domains, and generally, a
high diffraction efficiency is achieved only at rather large
depths, corresponding to aspect ratios that are not easily
realized in practice. Qualitatively, the four regimes described
above are a consequence of the effective refractive index
experienced by the fundamental mode in the grating region, and the
resulting reflectivities at the top and bottom boundary. The
effective refractive index of the fundamental mode is determined
by the mean refractive index. For large groove width, the
mean refractive index is near to unity. Thus, as the incident angle is
large, the fundamental mode in the grating region becomes
evanescent. In this case, a notable amount of light can traverse the
grating region only if the thickness of the grating is of the same
order as the penetration depth of the evanescent wave. This small
thickness conflicts with the interference condition for the two
modes in the grating region, that requires a sufficiently large
thickness to accumulate the necessary phase difference between the
fundamental and the first mode. Thus, for those groove widths, for
which the fundamental grating mode is evanescent, a high
diffraction efficiency cannot be achieved. For large incident angles
most of the reasonable choices for the width and depth
fall into this regime. This makes the design of a high dispersion
immersed grating a hard task.

It is remarkable, that also here, the introduction of a high index
material provides a solution, though, through a rather different
mechanism then observed previously with the TIR grating. While, in
case of the TIR grating, the high index material paved the way to
a grating with shallow grooves, by providing a high refractive
index contrast, here the high index material serves to increase
the mean refractive index in the grating region -- potentially
above the refractive index of glass -- to render the grating truly
transparent for any incident angle (for most of the possible
groove widths).

\begin{figure}
  \centering
  \includegraphics[scale=1]{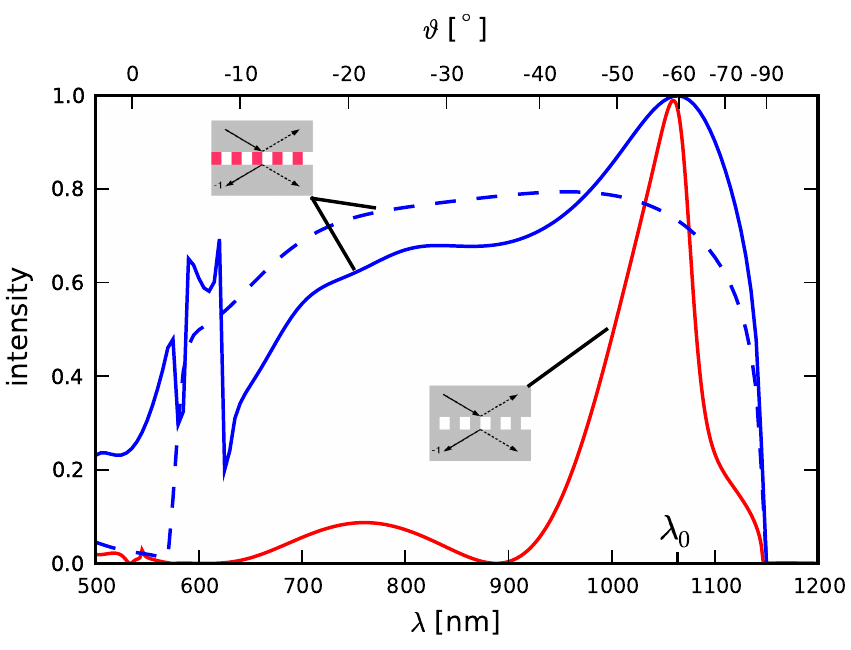}
  \caption{Spectral characteristics for an immersed dielectric grating
  featuring a high index material in the grating region (blue lines), as compared
  to its all-in-glass equivalent (red line). The solid blue line represents a
  design that is optimized for highest peak efficiency, corresponding
  to $w=0.43T$ and $d=1.3T$, as shown by the white square in
  Fig.\ref{fig:void-w-d}(b)). The dashed blue line represents a design
  that is optimized for maximum bandwidth, corresponding to $w=0.6T$
  and $d=0.7T$, as shown by the white diamond in Fig.\ref{fig:void-w-d}(b).}
  \label{fig:void-lambda}
\end{figure}

The four right panels of Fig.\ref{fig:void-w-d} shows the diffraction
efficiency of an immersed
transmission grating using a high index material in the
grating region (as before, we assume the refractive index of the high
index material is $n=2.4$). A region is opened up in the range
$w<0.6T$, where the
grating becomes transparent. Near 100\% diffraction
efficiency is achieved with
a number of possible choices of the width and
depth. Fig.\ref{fig:void-lambda} shows
the spectral characteristics for two possible choices of
the groove width and depth, as compared to the equivalent immersed
grating fabricated entirely from glass. The solid blue line corresponds to
$w=0.43T$ and $d=1.3T$, as shown by the white square in
Fig.\ref{fig:void-w-d}(b), and represents a
design that is optimized for highest peak efficiency in a narrow
band around the design wavelength. The dashed blue line corresponds to $w=0.6T$
and $d=0.7T$, as shown by the white diamond in
Fig.\ref{fig:void-w-d}(b), and represents a design
that is optimized for maximum bandwidth (around 80\% diffraction
efficiency over nearly the full octave).
To the best of our knowledge, the latter represents the largest
bandwidth transmission grating (with the large dispersion considered
here), devised so far.

\section{Large bandwidth and 100\% peak efficiency transmission
  grating in the classical scattering geometry}\label{sec:trans}

The successful improvement of the spectral and dispersive
characteristics of a dielectric TIR, as well as dielectric immersed
gratings, described in the preceding sections, intrigues the
question, whether a high index material can also contribute to improve
transmission gratings of the classical design.

Fig.\ref{fig:geometry}(c) shows the schematics of a classical
transmission grating. As in
the case of a dielectric TIR grating, light is diffracted at the
back side of a glass body. However, here the incident angle is smaller
than the angle of total internal reflection, such that the light is mainly
scattered into transmitted diffraction orders. This grating geometry
has a number of great advantages. (1) Due to refraction, the
transmitted orders leave the grating under a large angle, which, in
virtue of Eq.\ref{dispersion} corresponds to large dispersion. Because
of this, dielectric transmission gratings are typically superior over
classical reflection gratings in the large dispersion
regime (2) This is achieved without the need of a particularly small
period, since, in
virtue of Eq.\ref{period}, the period is determined by the incident
angle, and the latter is -- again, due to the refraction -- comparably
small. This facilitates fabrication in practice
(though, the period is not smaller than that of a gold grating with equal
dispersion -- it is the same).
(3) The diffracted wave lives in air, saving the necessity to
couple out the arc of diffracted rays through a flat or suitably
curved glass-air interface (in certain applications, such as
e.g. grism applications, the coupling
out through a glass-air interface as faced in case of the TIR
grating, may however also
be used to a benefit, e.g., its dispersive nature may be specifically
used to compensate for higher order dispersion \cite{gibson06ol31_3363}).

Because of these great advantages, the backside diffraction
geometry is to date the primary choice for most dielectric
grating applications.

\begin{figure}
  \centering
  \rput(-14mm,15mm){(a)}
  \hspace{-2mm}
  \includegraphics[scale=0.3]{icon-trans.pdf}
  \hspace{26mm}
  \rput(-12mm,15mm){(b)}
  \includegraphics[scale=0.3]{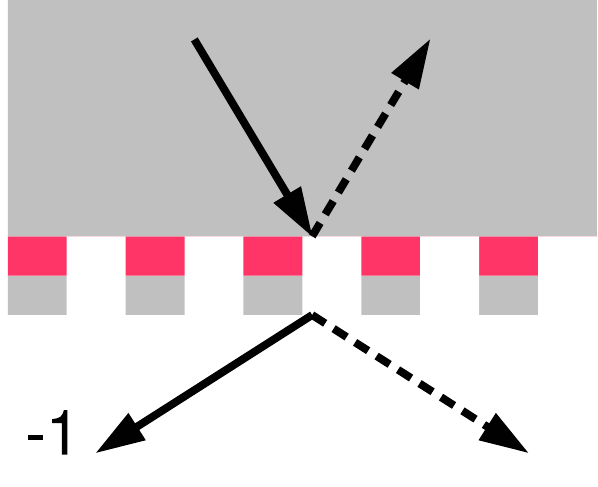}\\
  \includegraphics[scale=0.6]{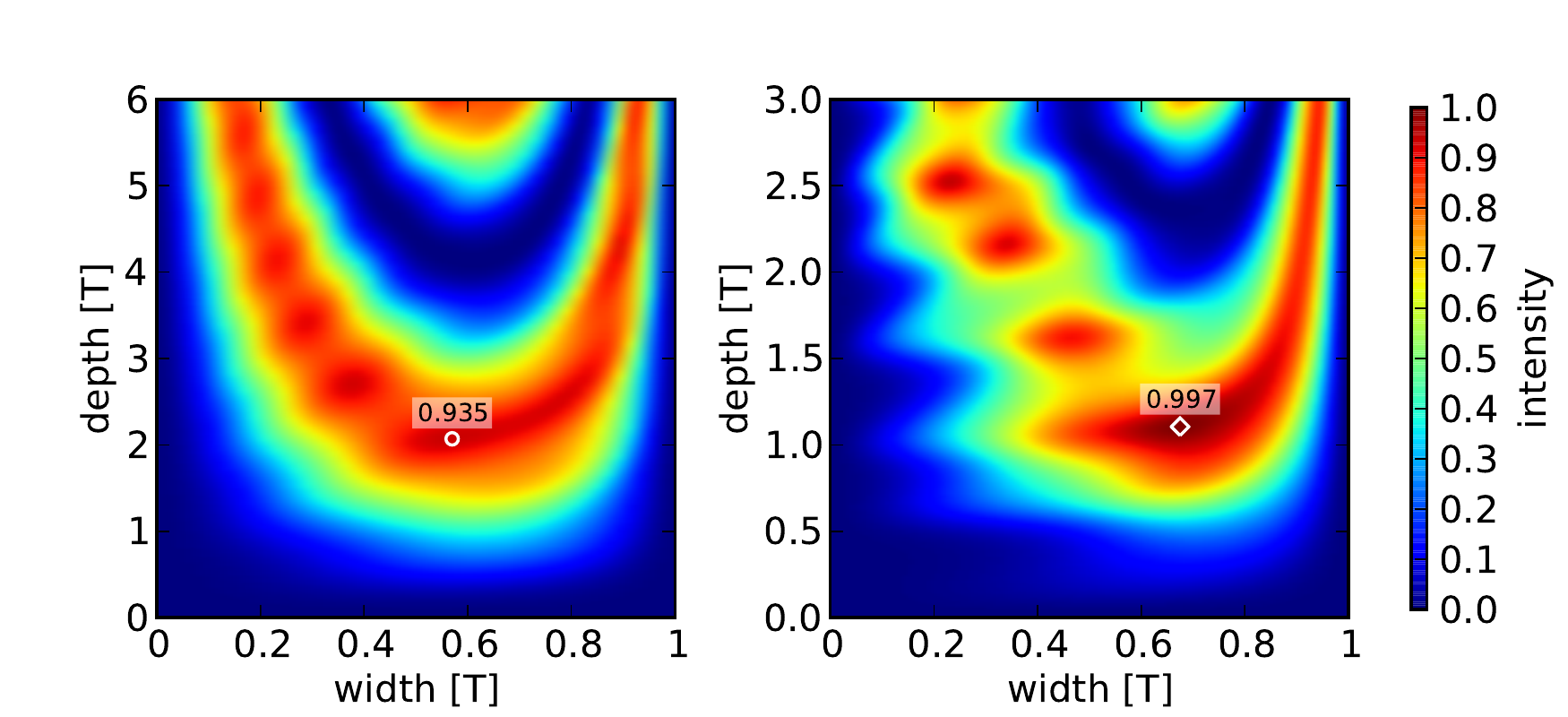}
  \caption{(a) diffraction efficiency of a
    classical dielectric transmission grating. (b) diffraction
    efficiency of a double-layer grating featuring a
    bured high index layer, with thickness ratio $r=0.5$. The circle
    and diamond mark those
    values of the groove width and depth, for which the spectral
    characteristics are evaluated in Fig.\ref{fig:trans-lambda}.}
  \label{fig:trans-w-d}
\end{figure}

\begin{figure}
  \centering
  \includegraphics[scale=1]{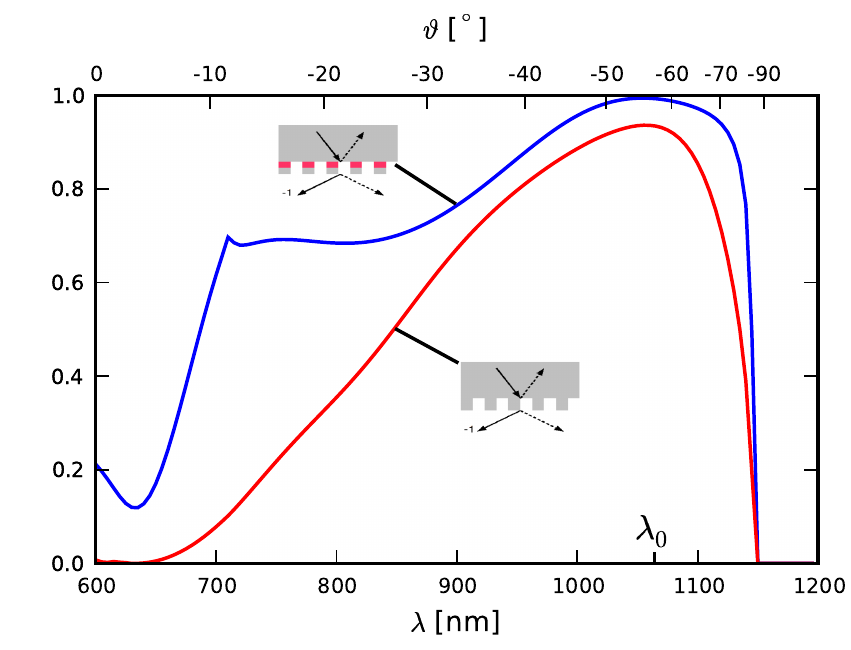}
  \caption{Spectral characteristics of a dielectric transmission grating
    featuring a buried high index layer (blue line), as compared to a
    classical transmission grating (red line). The groove width an
    depth are $w=0.68T$, $d=1.1T$, respectively
    $w=0.57T$, $d=2.07T$, as marked
    by the diamond and circle in Fig.\ref{fig:trans-w-d}.}
  \label{fig:trans-lambda}
\end{figure}

Fig.\ref{fig:trans-w-d}(a) shows the diffraction efficiency as a
function of groove width and depth, for a classical dielectric
transmission grating. The grating period $T=614.30$nm, and the
incident angle $\vartheta_0=36.67^\circ$ are chosen such that the 
diffraction angle of the transmitted $-1^{st}$ order is $60^\circ$,
such that both the diffraction angle and the angular dispersion are
the same as before. The maximum diffraction
efficiency is found with $w=0.57T$ and $d=2.07T$, on the center of the
first fringe, as 93.5\% (white circle).
The red line in Fig.\ref{fig:trans-lambda} shows the
corresponding spectral characteristics. The effective reflectivity of
the top and bottom surface of the grating region are not equal, such
that a reflection loss into the ordinary reflected order is
encountered, and the diffraction efficiency remains smaller than
100\%.

The qualitative understanding of the reflection loss suggests that
a low loss into the fundamental reflected order could be
achievable, if the effective refractive index in the grating
region for the fundamental mode was close to $\sqrt{n^R}$, and the
thickness of the grating region was close to an integer multiple of
$\lambda/(4n^R)$, as is the principle of single layer
anti-reflection coatings. In contrast, a large reflection loss
into the fundamental order was expected, if the effective
refractive index of the fundamental mode in the grating region was
high. A substitution of the glass in the grating region by a high
index material is therefore not promising. However it turns out
that a simple double-layer grating is already enough. To ensure a low
reflectivity at the bottom grating boundary, the high index grating
layer is buried underneath a glass layer, as shown in
Fig.\ref{fig:trans-w-d}(b). To find the optimum thickness of both
grating layers, we evaluated the diffraction efficiency as a
function of width, depth and ratio $r=d_h/d_g$ of the thickness
$d_h$ and $d_g$ of the high index layer, respectively the glass
layer (third supplemental movie). Fig.\ref{fig:trans-w-d}(b)
shows the optimum choice corresponding to $r=0.5$. The optimum
width and depth is found as $w=0.68T$ and $d=1.1T$. Remarkably, the
grating reaches effectively 100\% diffraction efficiency. Thus,
the TIR grating featuring a buried high index layer is indeed a
complementary approach to a transmission grating featuring 100\%
diffraction efficiency, and an alternative to the immersed
dielectric grating devised in \cite{clausnitzer08oe16_5577}.
Additionally, the spectral characteristics are greatly improved.
The blue line in Fig.\ref{fig:trans-lambda} shows the spectral
characteristics. The grating surpasses a classical transmission
grating over the entire spectrum. To the best of our knowledge, this
represents the largest bandwidth, high efficiency transmission
grating devised to date.

\section{Summary and conclusions}\label{sec:conclusions}
We have analyzed the diffraction characteristics of dielectric
gratings that feature a high index grating layer, and shown parameter
choices that are superior in terms of bandwidth and efficieny. A
qualitative understanding is supplied in combination with rigorous
calculations. The common grating types were considered,
including reflection gratings, immersed transmission gratings and
classical transmission gratings. Several profile types were explored,
including novel structures that
comprise a combination of enhanced material and enhanced geometry.
The suggested devices are within current
manufacturing capabilities. The gratings are compared to existing
technology and applications are pointed out. We expect these types
of gratings to become a useful addition to the existing range of
grating geometries.

\section*{Acknowledgements}
H.R. would like to thank Frieder Mugele for reserving him the freedom
and time that allowed him to do this work. We are grateful to Detlef
Lohse for providing us access to his computer cluster, on which most
of the calculations were performed.

\end{document}